\documentstyle[epsf,tighten,preprint,prd,aps]{revtex}

\newcommand{\Lie}{{\pounds}}
\renewcommand{\Re}{\text{Re}\,}
\renewcommand{\Im}{\text{Im}\,}

\begin{document}
\epsfverbosetrue
\draft


\title{
  Perturbations and Critical Behavior in the\\
  Self-Similar Gravitational Collapse of a Massless Scalar Field
}
\author{Andrei V. Frolov%
  \thanks{Email: \texttt{andrei@phys.ualberta.ca}}}
\address{
  Physics Department, University of Alberta\\
  Edmonton, Alberta, Canada, T6G 2J1}
\date{April 14, 1997}
\maketitle

\begin{abstract}
  This paper studies the perturbations of the continuously self-similar
  critical solution of the gravitational collapse of a massless scalar
  field (Roberts solution). The perturbation equations are derived and
  solved exactly. The perturbation spectrum is found to be not
  discrete, but occupying continuous region of the complex plane. The
  renormalization group calculation gives the value of the mass-scaling
  exponent $\beta=1$.
\end{abstract}

\narrowtext


\section{Introduction} \label{sec:intro}

Numerical calculation of spherically symmetric gravitational collapse
of a massless scalar field by Choptuik \cite{Choptuik:93} and
subsequent results for different matter models and symmetries (see for
example Refs. \cite{Abrahams&Evans:93,Evans&Coleman:94,%
Hirschmann&Eardley:95a,Hirschmann&Eardley:95b}) spectacularly
demonstrate that critical phenomena occur in the gravitational
collapse, and that the near-critical behavior is universal in some
important aspects.

There are two possible late-time outcomes of gravitational collapse,
distinguished by whether or not a black hole is formed in the course of
evolution. Which end state is realized depends on initial conditions,
specified by a control parameter $p$ that characterizes the
gravitational interaction strength in the ensuing evolution. For
$p<p^*$ (subcritical solution) the gravitational field is too weak to
form a black hole, while for $p>p^*$ (supercritical solution) a black
hole is produced. The critical solution, corresponding to a control
parameter value $p^*$ just at the threshold of black hole formation,
acts as an intermediate attractor for nearby initial conditions, and
has an additional symmetry: discrete or continuous self-similarity,
also referred to as echoing. For supercritical evolution there is
usually no mass gap, and the mass of the resultant black hole is
described by the mass-scaling relation
\begin{equation} \label{eq:scaling}
  M_{\text{BH}}(p) \propto |p-p^*|^\beta,
\end{equation}
where $\beta$ is mass-scaling exponent.

General gravitational collapse is a very difficult problem, even if
restricted to the case of spherical symmetry. As a result, most studies
of critical behavior resort to numerical calculations at some stage,
with few fully analytical solutions known. Finding critical solutions
is made easier by the fact that they have additional symmetry;
requiring solution to be continuously self-similar can simplify the
problem enough to make it solvable for simpler matter models. Both the
stability of the critical solution and the exact mass-scaling exponent
can then be determined by linear perturbation analysis, as suggested by
Evans and Coleman \cite{Evans&Coleman:94} and carried out in Refs.
\cite{Koike&Hara&Adachi:95,Hara&Koike&Adachi:96,Maison:96}.

In this paper we consider an analytical continuously self-similar
solution of gravitational collapse of a massless scalar field
constructed by Roberts \cite{Roberts:89}. Originally it was intended as
a counter-example to cosmic censorship conjecture, but it was later
rediscovered in context of the critical gravitational collapse by Brady
\cite{Brady:94} and Oshiro, Nakamura and Tomimatsu
\cite{Oshiro&Nakamura&Tomimatsu:94}. Although this solution does
exhibit critical behavior, there are two problems. First, as any
self-similar solution, it is not asymptotically flat, and the mass of
the black hole grows infinitely. This can be prevented if we cut the
self-similar solution away from the black hole and glue it to some
other solution to make the spacetime asymptotically flat and of finite
mass, which was in fact done in Ref. \cite{Wang&Oliveira:96}. A second,
and bigger, problem is that growing non-self-similar modes will
dominate the mass-scaling exponent if they exist, but they are not
accounted for by the self-similar solution.

This paper addresses the second problem by doing linear perturbation
analysis of the Roberts solution. We calculate the mass-scaling
exponent $\beta$ and analyze the stability of the critical solution.
The remarkable feature of this model is that it allows exact analytical
treatment.

\section{Self-Similar Solution} \label{sec:ss_soln}

We begin by presenting a continuously self-similar spherically
symmetric solution of the gravitational collapse of a massless scalar
field (Roberts solution). It is most easily derived in null
coordinates. Therefore we write the metric as
\begin{equation} \label{eq:metric}
  dS^2 = - 2e^{2\Sigma}\, du dv + R^2\, d\Omega^2,
\end{equation}
where $\Sigma$ and $R$ are functions of both $u$ and $v$. The
coordinates $u$, $v$ can be reparametrized as $u(\bar{u})$,
$v(\bar{v})$, so coordinate choice freedom is given by two functions of
one variable each; it can be fixed by setting $\Sigma=0$ at two null
hypersurfaces $v=\text{const}$ and $u=\text{const}$. With this choice
of metric the Einstein-scalar field equations become
\begin{mathletters} \label{eq:es}
\begin{equation} \label{eq:es:3}
  (R^2)_{,uv} + e^{2\Sigma} = 0,
\end{equation}
\begin{equation} \label{eq:es:4}
  2\Sigma_{,uv} - \frac{2R_{,u}R_{,v}}{R^2} -
    \frac{e^{2\Sigma}}{R^2} + 2 \Phi_{,u}\Phi_{,v} = 0,
\end{equation}
\begin{equation} \label{eq:es:5}
  R_{,vv} - 2\Sigma_{,v}R_{,v} + R (\Phi_{,v})^2 = 0,
\end{equation}
\begin{equation} \label{eq:es:6}
  R_{,uu} - 2\Sigma_{,u}R_{,u} + R (\Phi_{,u})^2 = 0,
\end{equation}
\begin{equation} \label{eq:es:7}
  \Phi_{,uv} + \frac{R_{,v}\Phi_{,u}}{R} + \frac{R_{,u}\Phi_{,v}}{R} = 0.
\end{equation}
\end{mathletters}
The first four equations (\ref{eq:es:3}-\ref{eq:es:6}) are the Einstein
equation for metric (\ref{eq:metric}), and equation (\ref{eq:es:7}) is
the scalar wave equation $\Box\Phi=0$. Equations (\ref{eq:es:5},%
\ref{eq:es:6}) are constraints. Note that in the case of a massless
scalar field the wave equation follows from the Einstein equations, so
one of the equations (\ref{eq:es:3},\ref{eq:es:4},\ref{eq:es:7}) is
redundant.

We turn on the influx of the scalar field at the advanced time $v=0$,
so that the spacetime is Minkowskian to the past of this surface. The
initial conditions for system (\ref{eq:es}) are specified there by the
continuity of the solution. We use coordinate freedom to set
$\Sigma|_{v=0}=0$.

Assumption that the collapse is continuously self-similar, i.e. that
there exists a vector field $\xi$ such that $\Lie_\xi g_{\mu\nu} =
2g_{\mu\nu}$, allows the Einstein-scalar equations (\ref{eq:es}) to be
solved analytically. The critical solution is given by
\begin{eqnarray} \label{eq:crit}
  \Sigma_0 &=& 0,\nonumber\\
  R_0 &=& \sqrt{u^2 - uv},\nonumber\\
  \Phi_0 &=& \frac{1}{2} \ln \left[1 - \frac{v}{u}\right],
\end{eqnarray}
and is a member of a one-parameter family of self-similar solutions
\begin{eqnarray} \label{eq:ss_soln}
  \Sigma &=& 0,\nonumber\\
  R &=& \sqrt{u^2 - uv - pv^2},\nonumber\\
  \Phi &=& \frac{1}{2} \ln
    \left[-\frac{2pv + u(1 - \sqrt{1+4p})}{2pv + u(1 + \sqrt{1+4p})}\right] +
    \varphi(p).
\end{eqnarray}
Above $\varphi(p)$ is constant component of a scalar field, chosen so
that $\Phi|_{v=0}=0$, and the solution converges to the critical
solution as $p$ goes to zero:
\begin{equation} \label{eq:const_phi}
  \varphi(p) =
    - \frac{1}{2} \ln \left[\frac{\sqrt{1+4p}-1}{\sqrt{1+4p}+1}\right] =
    - \frac{1}{2} \ln p + p + O(p^2).
\end{equation}
For values $p>0$ of control parameter, a black hole is formed; for
$p<0$ the field disperses, and the spacetime is flat again in region
$u>0$. (Our choice of sign of $p$ is opposite to the one in
\cite{Brady:94}. Also note that the constant field term is omitted
there.)

The existence and position of the apparent horizon is given by
condition $A(u,v)=0$, where
\begin{equation} \label{eq:ah}
  A(u,v) = u^{;\mu}_{\ \ ;\mu} = R_{,v}.
\end{equation}
Due to the spherical symmetry of the problem it is possible to
introduce a local mass $M(u,v)$ by
\begin{equation} \label{eq:mass}
  1 - \frac{2M}{R} = 2 g^{uv} R_{,u}R_{,v}.
\end{equation}
The mass contained inside the apparent horizon is the mass of the black
hole
\begin{equation} \label{eq:m_bh}
  M_{\text{BH}} = \frac{1}{2} R_{\text{AH}},
\end{equation}
where $R_{\text{AH}}$ is the radius of apparent horizon given by
equation (\ref{eq:ah}).

\section{Linear Perturbations} \label{sec:pert}

We now apply the perturbation formalism
\cite{Koike&Hara&Adachi:95,Hara&Koike&Adachi:96,Maison:96}. It is
convenient to analyze perturbations to the self-similar solution in new
variables natural to the problem. For this purpose we introduce a
``spatial'' coordinate $x$ and a ``scaling variable'' $s$
\begin{equation} \label{eq:xs}
  x = \frac{1}{2} \ln \left[1 - \frac{v}{u}\right], \ \ 
  s = - \ln(-u),
\end{equation}
with the inverse transformation
\begin{equation} \label{eq:uv}
  u = - e^{-s}, \ \ 
  v = e^{-s} (e^{2x} - 1).
\end{equation}
The signs are chosen to make the arguments of the logarithm positive in
the region of interest ($v>0$, $u<0$), where the field evolution occurs.

In these coordinates the metric (\ref{eq:metric}) becomes
\begin{equation} \label{eq:metric:xs}
  dS^2 = 2 e^{2(\Sigma + x - s)}
    \left[(1 - e^{-2x}) ds^{2} - 2 ds dx\right] +
    R^2\, d\Omega^2,
\end{equation}
and the critical solution (\ref{eq:crit}) is simply
\begin{equation} \label{eq:crit:xs}
  \Sigma_0 = 0, \ \ 
  R_0 = e^{x-s}, \ \ 
  \Phi_0 = x.
\end{equation}

We now consider small perturbations of critical solution
(\ref{eq:crit:xs}). Quite generally, the perturbation modes
can be written as
\begin{eqnarray} \label{eq:pert}
  \Sigma &=& \Sigma_0 + p \sigma(x) e^{ks},\nonumber\\
  R &=& R_0 + p r(x) e^{x-s} e^{ks},\nonumber\\
  \Phi &=& \Phi_0 + p \phi(x) e^{ks},
\end{eqnarray}
which amounts to doing Laplace transformation on general perturbations
dependent on both $x$ and $s$. It is understood that there can be many
perturbation modes, with distinct $\sigma$, $r$ and $\phi$ for each
(possibly complex) $k$ in the perturbation spectrum. To recover a
general perturbation we must take the sum of these modes. Modes with
$\Re k>0$ (called relevant modes) grow and eventually lead to black
hole formation, while modes with $\Re k<0$ decay and are irrelevant.

Let us discuss how the mass-scaling exponent $\beta$ is related to the
perturbation spectrum. The function $A$, defined by equation
(\ref{eq:ah}), takes the form
\begin{equation} \label{eq:pert:A}
  A(x,s,p) = A_0(x) - p a(x) e^{ks},
\end{equation}
where $A_0$ is the self-similar background term. Recall that $A$
vanishes at the apparent horizon. Even thought there is no apparent
horizon in critical solution, i.e. $A_0(x)>0$, the exponentially
growing perturbation will eventually make $A$ zero, leading to black
hole formation. The mode with the largest eigenvalue,
$\kappa=\max\{\Re k\}$, will clearly dominate, giving the position of
the apparent horizon $R_{\text{AH}}(x_{\text{AH}})$ by
\begin{equation} \label{eq:pert:ah}
  A_0(x_{\text{AH}}) = p a(x_{\text{AH}}) \,
    \frac{e^{\kappa x_{\text{AH}}}}{(R_{\text{AH}})^\kappa}.
\end{equation}
Here we made use of the fact that $R=e^{x-s}$ to the zeroth order in
$p$. Therefore the dependence of the black hole mass on $p$, taken to
be the control parameter, is
\begin{equation} \label{eq:pert:scaling}
  M_{\text{BH}} \propto R_{\text{AH}} \propto p^{1/\kappa}.
\end{equation}
This is precisely the mass-scaling relation (\ref{eq:scaling}) with
exponent $\beta=1/\kappa$. For a more detailed discussion of this
approach and its validity see Ref. \cite{Hara&Koike&Adachi:96}.

Perturbing the Einstein-scalar equations (\ref{eq:es}) by a mode
(\ref{eq:pert}), we obtain a system of linear partial differential
equations. After a change of variables $(u,v)$ to $(x,s)$, the
$s$-dependence separates, and we are left with a following system of
linear ordinary differential equations for $\sigma$, $r$, and $\phi$ in
the independent variable $x$:
\widetext
\begin{mathletters} \label{eq:p}
\begin{equation} \label{eq:p:3}
  (1 - e^{-2x}) r'' + 2 (k - e^{-2x}) r' + 4 (k - 1) r + 4 \sigma = 0,
\end{equation}
\begin{equation} \label{eq:p:4}
  (1 - e^{-2x}) \sigma'' + 2 (k + e^{-2x}) \sigma' - 4 \sigma
  + 2 (1 - e^{-2x}) \phi' + 2k \phi
  + 2 e^{-2x} r' + 2 (2-k) r = 0,
\end{equation}
\begin{equation} \label{eq:p:5}
  r'' - 2 \sigma' + 2 \phi' = 0,
\end{equation}
\begin{eqnarray} \label{eq:p:6}
  &(e^{2x} - 2 + e^{-2x}) r'' - 4k (1 - e^{2x}) r' - 4k (1 - k e^{2x}) r
  + 2 (e^{2x} - e^{-2x}) \sigma' + 4k (1 + e^{2x}) \sigma \nonumber\\
  &+ 2 (e^{2x} - 2 + e^{-2x}) \phi' - 4k (1 - e^{2x}) \phi = 0,
\end{eqnarray}
\begin{equation} \label{eq:p:7}
  (1 - e^{-2x}) \phi'' + 2k \phi' + 2k \phi + 2 (1 - e^{-2x}) r' + 2k r = 0.
\end{equation}
\end{mathletters}
\narrowtext
Here a prime denotes the derivative with respect to $x$.

Boundary conditions for the system (\ref{eq:p}) are specified at $x=0$
and $x=\infty$. We require that perturbations grow slower than
background at large $x$ (i.e. that $\sigma$ and $r$ are bounded, and
$\phi$ grows slower than $x$) for the perturbation expansion to be
valid. Because the solution must be continuous at the hypersurface
$v=0$, we have $r(0)=\phi(0)=0$. Since we used coordinate freedom in
$u$ to set $\Sigma$ to zero on that surface, we also have
$\sigma(0)=0$. Thus, the boundary conditions are
\begin{eqnarray} \label{eq:bc}
  &\sigma(0) = r(0) = \phi(0) = 0, \nonumber\\
  &\sigma(x), r(x), \phi(x) \text{ grow slowly as } x \rightarrow \infty.
\end{eqnarray}
Equations (\ref{eq:p}) together with boundary conditions (\ref{eq:bc})
constitute our eigenvalue problem.

The infinitesimal coordinate transformation $v \mapsto v - 2\zeta(v)$
corresponding to the remaining coordinate freedom in $v$ will give rise
to unphysical gauge modes
\begin{eqnarray} \label{eq:gauge}
  \delta\Sigma &=& \zeta'(v)
    = \zeta'[e^{-s}(e^{2x}-1)], \nonumber\\
  \delta R &=& \frac{\zeta(v)}{(1-v/u)^{1/2}}
    = e^{-x} \zeta[e^{-s}(e^{2x}-1)], \nonumber\\
  \delta\Phi &=& \frac{\zeta(v)}{(u-v)}
    = - e^{s-2x} \zeta[e^{-s}(e^{2x}-1)].
\end{eqnarray}
Since on a $v=0$ hypersurface $\delta\Sigma$, $\delta R$, and
$\delta\Phi$ must vanish identically, $\zeta(0)=\zeta'(0)=0$, so a
Taylor expansion of $\zeta(v)$ around zero starts from $v^2$ term.
Therefore these gauge modes correspond to negative eigenvalues $k$, and
are irrelevant.

\section{Solution of Perturbation Equations} \label{sec:pert_soln}

Observe that there is an obvious solution of perturbation equations
(\ref{eq:p}), which can be obtained from self-similar solution
(\ref{eq:ss_soln}) by expanding around $p=0$
\begin{eqnarray} \label{eq:ss_mode}
  \Sigma &=& 0,\nonumber\\
  R &=& \sqrt{u^2 - uv} \left(1 - \frac{p}{2}\, \frac{v^2}{u^2 - uv}\right)
    = e^{x-s} \left(1 - 2 p \sinh^2 x\right),\nonumber\\
  \Phi &=& \frac{1}{2} \ln \left[1 - \frac{v}{u}\right] +
    \frac{p}{2}\, \frac{v^2 - 2uv}{u^2 - uv}
    = x + p \sinh 2x.
\end{eqnarray}
Clearly, the self-similar mode has eigenvalue $k=0$, hence any relevant
mode will dominate the calculation of mass-scaling exponent.

The perturbation equations (\ref{eq:p}) inherit their structure from
general Einstein-scalar equations (\ref{eq:es}); equations
(\ref{eq:p:3},\ref{eq:p:4},\ref{eq:p:7}) are dynamical equations, while
equations (\ref{eq:p:5},\ref{eq:p:6}) are constraints. Equation
(\ref{eq:p:4}) is redundant
\begin{equation}
  (\ref{eq:p:4}) = [(\ref{eq:p:3})' - 2 (\ref{eq:p:3}) 
  - (1 - e^{-2x}) (\ref{eq:p:5})' - 2 (e^{-2x} + k - 1) (\ref{eq:p:5})
  + 2 (\ref{eq:p:7})]/2,
\end{equation}
so it will be discarded. The simple integrable form of equation
(\ref{eq:p:5}) allows elimination of one of the unknowns, say $\sigma$,
from the other equations by
\begin{equation} \label{eq:no_sigma}
  \sigma = r'/2 + \phi + C.
\end{equation}
Thus we are left with two second order ODE's for $r$ and $\phi$:
\begin{mathletters} \label{eq:pe}
\begin{equation} \label{eq:pe:1}
  (1 - e^{-2x}) r'' + 2 (k + 1 - e^{-2x}) r' + 4 (k - 1) r + 4 \phi + 4C = 0,
\end{equation}
\begin{equation} \label{eq:pe:2}
  (1 - e^{-2x}) \phi'' + 2k \phi' + 2k \phi + 2 (1 - e^{-2x}) r' + 2k r  = 0
\end{equation}
with boundary conditions (\ref{eq:bc}), and one constraint
(\ref{eq:p:6}), which we equivalently rewrite as first order ODE with
the use of equation (\ref{eq:p:3})
\begin{eqnarray} \label{eq:pe:c}
 &(k - 2)(1 - e^{-2x}) r' + 2 (k^2 - k e^{-2x} - 2k + 2) r \nonumber\\
 & + 2 (1 - e^{-2x}) \phi' + 4 (k - 1) \phi + 2C (k + k e^{-2x} - 2) = 0.
\end{eqnarray}
\end{mathletters}
Equations (\ref{eq:pe:1},\ref{eq:pe:2}) can be separated with
introduction of the auxiliary function $z=r-\phi$, reducing to
\begin{mathletters} \label{eq:se}
\begin{equation} \label{eq:se:1}
  (1 - e^{-2x}) z'' + 2 k z' + 2 (k-2) z = -4 C,
\end{equation}
\begin{equation} \label{eq:se:2}
  (1 - e^{-2x}) r'' + 2 (k + 1 - e^{-2x}) r' + 4k r = 4 (z - C).
\end{equation}
\end{mathletters}
We convert this into algebraic form using change of variable
\begin{equation} \label{eq:y}
  y = e^{2x}, \ \ 
  x = \frac{1}{2} \ln y.
\end{equation}
Finally, equations (\ref{eq:se}) and constraint (\ref{eq:pe:c}) become
\begin{mathletters} \label{eq:ae}
\begin{equation} \label{eq:ae:1}
  y(1-y)\, \frac{d^2z}{dy^2} + [1 - (k+1)y]\, \frac{dz}{dy} - (k/2 - 1) z = C,
\end{equation}
\begin{equation} \label{eq:ae:2}
  y(1-y)\, \frac{d^2r}{dy^2} + [2 - (k+2)y]\, \frac{dr}{dy} - k r = C - z,
\end{equation}
\begin{equation} \label{eq:ae:c}
  2 y(1-y)\, \frac{dz}{dy} - 2 y(k-1) z
    - k y(1-y)\, \frac{dr}{dy} - k(1-ky) r + C [k + (k-2)y] = 0,
\end{equation}
\end{mathletters}
and boundary conditions (\ref{eq:bc}) are specified at $y=1$ and
$y=\infty$ by
\begin{eqnarray} \label{eq:abc}
  z(1)=r(1)=0,\ \ \frac{dr}{dy}(1) + C = 0, \nonumber\\
  z(y), r(y) \text{ grow slowly as } y \rightarrow \infty.
\end{eqnarray}
``Grow slowly'' means that $r(y)$ must be bounded, and $z(y)$ grows
at most logarithmically, with $y\, dr/dy + r - z$ bounded, as
$y \rightarrow \infty$. The last condition follows from $\sigma$ being
bounded at infinity.

Equation (\ref{eq:ae:c}) is indeed a constraint, since
$d(\ref{eq:ae:c})/dy = (\ref{eq:ae:1}) - k (\ref{eq:ae:2})$,
and is automatically satisfied for all $y$ if
\begin{equation} \label{eq:constr}
  (k-1) [k r(1) - 2 z(1) + 2 C] = 0
\end{equation}
is satisfied initially at $y = 1$. Imposing initial conditions
(\ref{eq:abc}), the constraint (\ref{eq:constr}) yields
\begin{equation} \label{eq:ck}
  C(k-1)=0,
\end{equation}
i.e. $C=0$ unless $k=1$.

Observe that equations (\ref{eq:ae:1}) and (\ref{eq:ae:2}) are
inhomogeneous hypergeometric equations. Their properties are well-known
and described in a number of books on differential equations (see, for
example, \cite{Erdelyi:53}), so the system (\ref{eq:ae}) can be
analytically solved and the perturbation spectrum exactly determined.

Equation (\ref{eq:ae:1}) is the homogeneous (except when $k=1$)
hypergeometric equation
\begin{equation} \label{eq:hyper}
  y(1-y)\, \frac{d^2z}{dr^2} + [c - (a+b+1)y]\, \frac{dz}{dy} - ab z = 0
\end{equation}
with coefficients
\begin{eqnarray} \label{eq:coeff:1}
  &c = 1,\ \ a+b = k,\ \ ab = k/2 - 1, \nonumber\\
  &a = 1/2\, (k - \sqrt{k^2-2k+4}),\ \ b = 1/2\, (k + \sqrt{k^2-2k+4}).
\end{eqnarray}
It has singular points at $y=0,1,\infty$, and its general solution is a
linear combination of any two different solutions from the set
\begin{eqnarray} \label{eq:gen:z}
  z_1 &=& F(a, b; k; 1-y),\nonumber\\
  z_2 &=& (1-y)^{1-k} F(1-a, 1-b; 2-k; 1-y),\nonumber\\
  z_3 &=& (-y)^{-a} F(a, a; a+1-b; y^{-1}),\nonumber\\
  z_4 &=& (-y)^{-b} F(b, b; b+1-a; y^{-1}),
\end{eqnarray}
where $F(a, b; c; y)$ is the hypergeometric function; $F(a,b;c;0) = 1$.
Any three of the functions (\ref{eq:gen:z}) are linearly dependent with
constant coefficients, for example
\begin{equation} \label{eq:hyper:connect}
  z_2 =
    \frac{\Gamma(2-k)\Gamma(b-a)}{\Gamma(1-a)^2}
      e^{-i\pi(1-b)} z_3 +
    \frac{\Gamma(2-k)\Gamma(a-b)}{\Gamma(1-b)^2}
      e^{-i\pi(1-a)} z_4.
\end{equation}
Functions $z_1,z_2$ and $z_3,z_4$ have the asymptotic behavior
\begin{eqnarray} \label{eq:asympt:z}
  z_1 \simeq 1,\ 
  z_2 \simeq (1-y)^{1-k} &\ \ \text{near}\ \ & y=1,\nonumber\\
  z_3 \simeq (-y)^{-a},\ 
  z_4 \simeq (-y)^{-b} &\text{near}& y=\infty.
\end{eqnarray}
The case when $k=1$ is special. In this case the solution
(\ref{eq:gen:z}) degenerates and $z_1 = z_2$ identically. In order to
deal with this situation we continue to use $z_1$ as earlier, but
denote by $z_2$ a solution independent of $z_1$. It is easy to verify
that $z_2$ is logarithmically divergent at $y=1$. Note also that for
$k=1$ we can have $C \ne 0$, so that the equation (\ref{eq:ae:1}) is
inhomogeneous, and we must add the particular solution $z=C$ to the
general solution above.

\begin{figure}
\centerline{
  \epsfxsize=\columnwidth \multiply\epsfxsize 4 \divide\epsfxsize 5
  \epsfbox{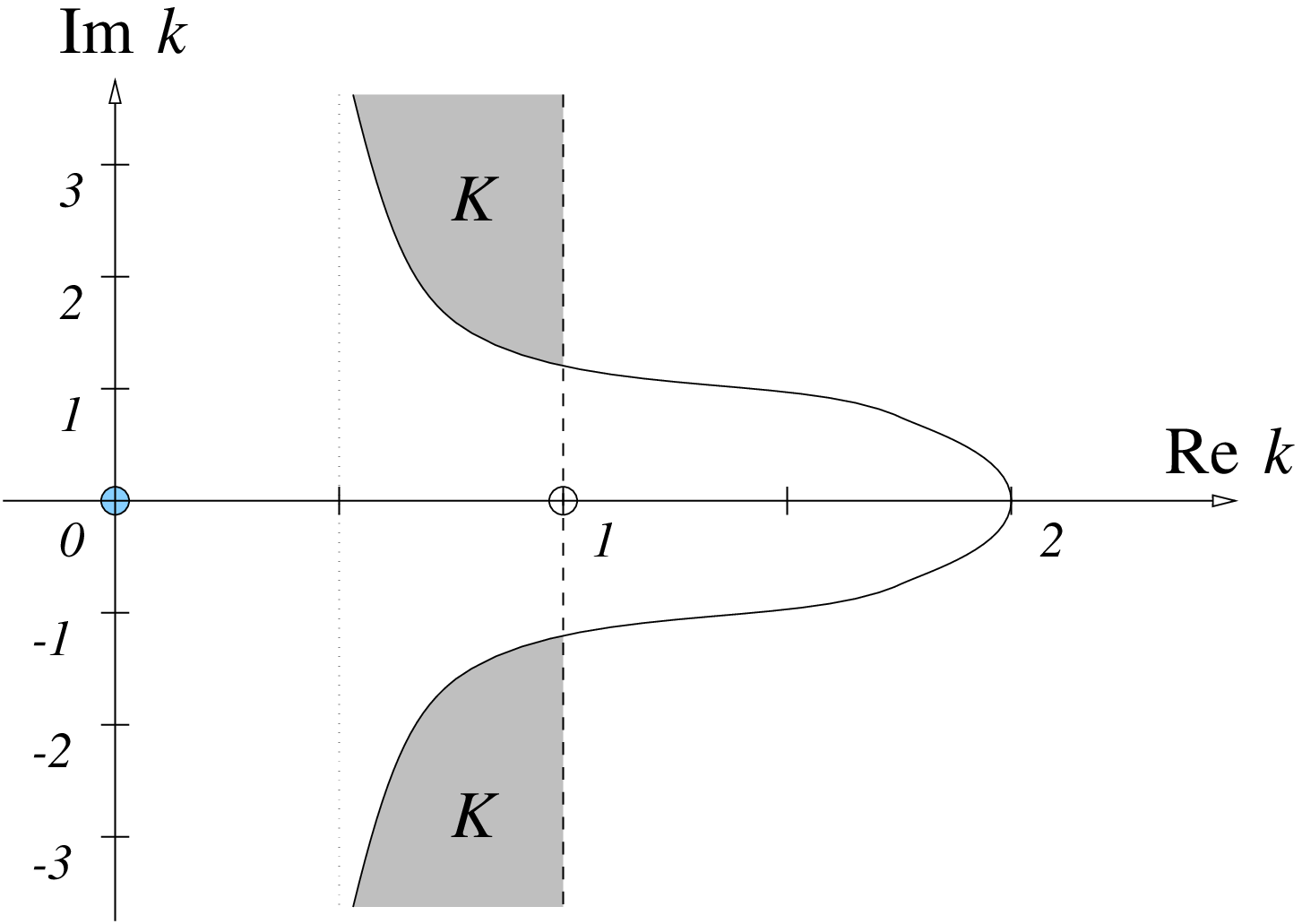}
}
\caption{
  Complex perturbation spectrum. Values of $k$ to the left of the solid
  line are prohibited by the boundary conditions at infinity, to the
  right of the broken line by the initial conditions at $y=1$. Values
  in the region of intersection (the shaded region $K$) are allowed,
  and constitute the perturbation spectrum.
} 
\label{fig:spectrum}
\end{figure}

Observe that at $y=1$, $z_2$ diverges as a power for $\Re k>1$,
logarithmically for $k=1$, does not have a limit for
$\Re k=1,\ \Im k\ne0$, and converges (to zero) only for $\Re k<1$. The
solutions that satisfy the initial condition $z(1)=0$ are
$z=c z_2,\ \Re k<1$ and $z=C(1-z_1),\ k=1$. From equation
(\ref{eq:hyper:connect}) we see that $z_2$ (and $z_1$) is connected to
$z_3$, $z_4$ by a linear relation with non-zero coefficients, so the
boundary conditions at infinity will only be satisfied if both $z_3$,
$z_4$ do not blow up, i.e. if $\Re a>0$, $\Re b>0$. The curve
$\Re a=0$, written in terms the real and imaginary parts of $k$, has
the form
\begin{equation} \label{eq:sep}
  (\Im k)^2 = \frac{\Re k\, (2 - \Re k)}{1 - (2\Re k)^{-1}},
\end{equation}
which divides the complex $k$-plane into two regions shown in Fig.
\ref{fig:spectrum}; to the left of the curve $\Re a<0$, and to the
right $\Re a>0$. So to satisfy all boundary conditions (\ref{eq:abc}),
$k$ must lie in the shaded region $K$ of the complex plane, and for any
$k \in K$,
\begin{equation} \label{eq:soln:z}
  z = (1-y)^{1-k} F(1-a, 1-b; 2-k; 1-y)
\end{equation}
is a solution of equation (\ref{eq:ae:1}) with boundary conditions
(\ref{eq:abc}).

Equation (\ref{eq:ae:2}) is also hypergeometric, with coefficients
\begin{equation} \label{eq:coeff:2}
  c = 2,\ \ a = 1,\ \ b = k.
\end{equation}
The general solution of the homogeneous equation is
\begin{equation} \label{eq:gen:r}
  r_1 = y^{-1},\ \ r_2 = y^{-1} (1-y)^{1-k},
\end{equation}
and the solution of inhomogeneous equation is easily constructed from
(\ref{eq:gen:r}) by
\begin{equation} \label{eq:soln:r}
  r = - \frac{1}{y(1-k)} \left[
    \int\limits_1^y z(\tilde{y}) d\tilde{y} -
    (1-y)^{1-k} \int\limits_1^y
      \frac{z(\tilde{y}) d\tilde{y}}{(1-\tilde{y})^{1-k}}
    \right].
\end{equation}
The limits of integration are chosen so that the boundary conditions at
$y=1$ are satisfied. Asymptotic (\ref{eq:asympt:z}) and equation
(\ref{eq:soln:r}) give the behavior
\begin{equation} \label{eq:asympt:r}
  r \simeq - \frac{y^{-1} (1-y)^{2-k}}{2-k}\ \ \text{near}\ \ y=1.
\end{equation}
Clearly, $r(1)=dr/dy(1)=0$. Also, $r$ is bounded at infinity since $z$
is, for the constraint (\ref{eq:ae:c}) requires
$k\, dr/dy(\infty) - 2\, dz/dy(\infty) = 0$. Therefore,
(\ref{eq:soln:z}) and (\ref{eq:soln:r}) for $k \in K$ is a solution of
perturbation equations (\ref{eq:ae}) with boundary conditions
(\ref{eq:abc}), and the region $K$ is the perturbation spectrum, which
turns out to be continuous.

\section{Conclusion} \label{sec:conclusion}

We have perturbed the continuously self-similar critical solution of
the gravitational collapse of a massless scalar field (Roberts
solution), and solved the perturbation equations exactly. The
perturbation spectrum was found to be not discrete, but continuous, and
occupying the region of the complex plane
\begin{equation}
  \frac{1}{2} < \Re k < 1,\ \ 
  |\Im k| > \sqrt{\frac{\Re k\, (2 - \Re k)}{1 - (2\Re k)^{-1}}},
\end{equation}
which, according to \cite{Hara&Koike&Adachi:96}, suggests
non-universality of critical behavior for different ingoing
wavepackets. The complex oscillatory modes might lead to decay of a
continuously self-similar solution (\ref{eq:crit}) into discrete
self-similar choptuon observed in \cite{Choptuik:93}.

The eigenvalues $k$ approach $\sup \Re k = 1$, which corresponds to the
mass-scaling exponent $\beta=1$. This is different from the exponent
$\beta=1/2$ found in \cite{Brady:94,Oshiro&Nakamura&Tomimatsu:94,%
Wang&Oliveira:96} using the self-similar solution.

\section*{Acknowledgments}

This research was supported by Natural Sciences and Engineering
Research Council of Canada. I am grateful to D. N. Page for interest
in this work and stimulating discussions.




\end{document}